\magnification=1200
\settabs 18 \columns
\baselineskip=17 pt
\font\stepone=cmb10 scaled\magstep1
\font\steptwo=cmb10 scaled\magstep2

\def\b{\bigskip}
\def\bb{\bigskip\bigskip}

\def\sqr#1#2{{\vcenter{\vbox{\hrule height.#2pt
 \hbox{\vrule width.#2pt height#1pt \kern#1pt
 \vrule width.#2pt} \hrule height.#2pt}}}}
\def\square{\mathchoice \sqr65 \sqr65 \sqr{2.1}3 \sqr{1.5}3}

\def\operp{\hbox{${\kern+.25em{\bigcirc}
\kern-.85em\bot\kern+.85em\kern-.25em}$}}
\def\lsim{\;\raise0.3ex\hbox{$<$\kern-0.75em\raise-1.1ex\hbox{$\sim$}}\;}
\def\gsim{\;\raise0.3ex\hbox{$>$\kern-0.75em\raise-1.1ex\hbox{$\sim$}}\;}
\def\no{\noindent}
\def\r{\rightline}
\def\ce{\centerline}
\def\ve{\vfill\eject}
\def\rdots{\mathinner{\mkern1mu\raise1pt\vbox{\kern7pt\hbox{.}}\mkern2mu
 \raise4pt\hbox{.}\mkern2mu\raise7pt\hbox{.}\mkern1mu}}

\def\e e{$e^+ e^-$ }

\def\ph{\varphi}
\def\onetwo{\raise.5mm\hbox{{${\scriptstyle{1\over 2}}$}}}
\def\3/2{\raise.5mm\hbox{{${\scriptstyle{3\over 2}}$}}}
\def\five/4{\raise.5mm\hbox{{${\scriptstyle{5\over 4}}$}}}
\def\fivetwo{\raise.5mm\hbox{{${\scriptstyle{5\over 2}}$}}}
\def\onethree{\raise.5mm\hbox{{${\scriptstyle{1\over 3}}$}}}
\def\one/4{\raise.5mm\hbox{{${\scriptstyle{1\over 4}}$}}}
\def\itwo{\raise.5mm\hbox{{${\scriptstyle{i\over 2}}$}}}
\def\to{\rightarrow}
\ve
\r {UCLA/98/TEP/8}
\r {March 1998}
\bb\bb\bb\bb
\ce{\steptwo  INTERACTING SINGLETONS}
\bb
\ce {Mosh\'e Flato}

\b
\ce{\it Physique Math\'ematique, Universit\'e de Bourgogne, Dijon, France}
\b
\ce {Christian Fr\o nsdal}
\b
\ce {\it Physics Department, University of California, Los Angeles, CA
90095-1547}
\bb\bb

\no {\it ABSTRACT.} There is a chance that singleton fields,
that in the context of strings and membranes have been regarded as
topological gauge fields that can interact only at
the boundary  of anti-De Sitter space, at spatial infinity, may have a more
physical manifestation
as costituents of massless fields in space time. The composite character of
massless fields is
expressed by field - current
identities that relate ordinary massless field operators to singleton
currents and stress-energy tensors. Naive versions of
such identities do not make sense, but when the singletons are described in
terms of dipole
structures, then such constructions are at least formally possible. The new
proposal includes and generalizes an early
composite version of QED, and includes quantum gravity, super gravity and
models of QCD. Unitarity of such theories is conjectural.

\ve

\no{\stepone 1. Introduction.}

 Recent developments in supergravity and string/membrane theory [1] point
to a form of duality
between massless fields on anti-De Sitter space (``the bulk") and
conformal field theory on the
boundary. The boundary values of bulk massless fields have all the quantum
numbers of composite
operators of the boundary conformal field theory [2], and it is tempting to
identify these two classes
of objects with each other. The fields of the boundary conformal field
theory are  the boundary
values of singleton fields in the bulk; therefore, the next step would be
to identify the massless
fields in the bulk with local bilinears in the singleton fields.

 This was already attempted long ago, not in five dimensions, where the
recent activity is taken place, but in ordinary,
4-dimensional space time. The result was a genuine composite version of QED
[3]. It was also
suggested, but merely suggested, that this idea may be applicable to
gravity and to the strong
interactions, that in some sense quarks are singletons [4]. That would, at
least, account for their
not being observed.

The recent developments [1] point to a concept of field-current identity
that could, with some luck,
be applied to construct a composite field model of massless fields in
general, including photons,
gravitons and perhaps gluons. Let $J$ be a conserved current; for example,
the usual vector current of a
conventional scalar, complex field. Let $\varphi$ be a complex, spinless
singleton field, and
introduce the interaction
$$
g^2\int d^4xJ^\mu (\bar\varphi\partial_\mu \varphi).\eqno(1.1)
$$
This is not manifestly  gauge invariant.

A singleton gauge transformation is a shift of $\varphi$ by a field
$\lambda$ that is perfectly general except that it falls off,
at spatial infinity, faster than the physical modes of $\varphi$.  The
strong association between gauge invariance and unitarity
suggests that any acceptable interaction must be insensitive to gauge
fields and that, consequently, only the boundary
values of the singleton field at infinity can participate [5].

Here we want to suggest that there are ways to circumvent this difficulty.
It was shown in [4,6] that
the interaction (5.1) can be made gauge invariant in quantum theory by an
alternative form of field
quantization. The effect of modifying the free field commutation relations
is that the classically
trivial interaction $g\int d^4 x J^\mu\partial_\mu \varphi$ (here $\varphi$
is real) gives rise to
an effective interaction of the form
$$
g^2\int d^4xJ^\mu :(\varphi\partial_\mu \varphi):,\eqno(1.2)
$$
where the colons stand for a kind of normal ordered product. In fact, this
normal ordered product could be identified with the
electromagnetic potential (quantum field operator).

Our new proposal is different. We suggest, in effect, that the lack of
gauge invariance of (1.1) may
be less destructive of unitarity than it appears at first sight. If this
proves to be the case, then
it may become possible to extend the field-current identity to the bulk;
that is, it would then be
feasible to interpret the quantity
$$
A_\mu = \bar\varphi  \partial_\mu \varphi\eqno(1.3)
$$
as the electromagnetic potential, even classically.

There is some evidence that suggests that the lack of gauge invariance of
(1.1)  may be of a benign
sort. If both factors,
$\bar\varphi$ and $\varphi$, are on shell, then the contribution of gauge
modes to the field (1.3) is a
gradient! This was  shown in [6]. In this paper we do not present a
definitive analysis of the
problem of unitarity. We hope to be able to do so in the future. The last
section makes some
suggestions.

Let us be more precise concerning the composite operators that seem to be
related to
the boundary values of massless fields: they are precisely the conserved
currents of the boundary
conformal field theory. We expect this to be true in the bulk as well, and
in fact, Eq.(1.3)
identifies the massless potential with the vector current of a spinless
field, usually conserved.
Similarly, the gravitational potential can be expected to be related to the
energy-momentum tensor
of the singleton field.

But this naive identification is not possible in quantum field theory, if
the current is conserved. For it is well known that
the quantum field operator, the potential, is not divergenceless.
In fact it is divergenceless
only when projected on the physical subspace defined by the Lorentz condition.
So what is needed is a
current that is conserved on this physical subspace only.

Such currents (and energy-momentum tensors) are in fact characteristic of
singleton field theories. The field equation for a
scalar singleton field, in quantum field theory, is
$$
(\square + u)\varphi = b, ~~(\square + u)b = 0,\eqno(1.4)
$$
where u is a constant, $b$ is the Nakanishi-Lautrup field and $\square$ is
the covariant
d'Alembertian. In this theory the conserved current is of the form
$\bar{\ph}\partial_\mu b$;
it vanishes on the physical subspace defined by the physicality (Lorentz)
condition, which in this
theory is $b^+|...> = 0$. We set
$$
\hat j^\mu = \sqrt{-g}g^{\mu\nu}(\bar\ph\partial_\nu \varphi -
(\partial_\nu\bar\ph)\varphi),~~\varphi_\mu :=\partial_\mu\varphi, \eqno(1.5)
$$
where $(g_{\mu\nu})$ is the anti-De Sitter metric, and find that
$$
\partial_\mu \hat j^\mu = \bar\ph b - \bar b\varphi.\eqno(1.6)
$$
The divergence of this current, but not the current itself, vanishes on the
physical subspace.  A
field current identification of the form (1.3), between quantum field
operators,
$$
A_\mu \propto \hat j_\mu,
$$
is therefore a possiblity. We learn that the constituents of massless
particles have to be described by gauge fields.

The conserved energy momentum tensor of a real, scalar singleton field is
$$
t^{\mu\nu} = \onetwo \sqrt{-g}g^{\mu\sigma}g^{\nu\tau}\bigl(\varphi_\sigma
b_\tau + b_\sigma\ph_\tau-
 g^{\lambda\rho}\varphi_\lambda b_\rho + u\,\varphi b\bigr).\eqno(1.7)
$$
We define
$$
\hat t^{\mu\nu} := \sqrt{-g}g^{\mu\sigma}g^{\nu\tau}\bigl(\varphi_\sigma
\varphi_\tau -
\onetwo(g^{\lambda\rho}\varphi_\lambda\varphi_\rho -
u\,\varphi\varphi)\bigr).\eqno(1.8)
$$
It satisfies
$$
\partial_\mu \hat t^{\mu\nu} =  \sqrt{-g}g^{\nu\sigma}
b\,\varphi_\sigma,\eqno(1.9)
$$
and it makes sense to suggest that
$$
h_{\mu\nu} \propto \hat t_{\mu\nu},
$$
with $(h_{\mu\nu})$ the gravitational potential (a perturbation of the
anti-De Sitter metric).
\b
\no{\it Outline.}

Section 2 deals with the scalar singleton field and shows in some detail
the basis for the construction
summarized above. Sections 3 and 4 introduce the spinor singleton and the
super singleton. It is shown, but
with less attention to the details, that the super singleton is the natural
object with which to
construct a version of super gravity in which all the massless particles
are singleton composites.
In Section 5 we make some additional remarks about the problem of unitarity
of these theories.

\ve

\no{\stepone 2. The scalar singleton.}
\b
\no{\it Vector potential.}

The scalar singleton is a scalar field that satisfies the dipole equation [6]
$$
(\square + u)^2\varphi = 0, ~~ u = -\five/4 \rho,\eqno(2.1)
$$
where $\rho$ is the anti-De Sitter curvature constant, henceforth set equal
to 1. The second order of the Klein-Gordon operator is
required in order that the propagator contain the modes of the singleton
representation
$
D(\onetwo,0)
$
of the anti-De Sitter group. The ordinary wave equation $(\square +
u)\varphi = 0$ is appropriate for $D(\fivetwo,0$). Actually, the
space of solutions of the dipole equation carries the following
Gupta-Bleuler triplet,
$$
D(\fivetwo,0) \rightarrow D(\onetwo,0) \rightarrow D(\fivetwo,0),\eqno(2.2)
$$
including physical modes (center), gauge modes (on the right) and their
canonical conjugates
(on the left). It is convenient to introduce a Nakanishi-Lautrup field $b$,
then the dipole equation takes the form (1.3).
The complete action, including Faddeev-Popov ghost $c$ and anti-ghost $d$
is [6]
\def\L{{\cal L}}
\def\ph{\varphi}
$$\eqalign{
\L &= \L_4 + \L_3,\cr
\L_4 &= \int d^4x\,\sqrt{-g}\bigl(g^{\mu\nu}\bar\ph_\mu b_\mu - u\,\bar \ph
b + \bar bb
- g^{\mu\nu}\bar c_\mu d_\nu + u\,\bar cd\bigr) + {\rm conj.})\cr
\L_3 &=\onethree\int d^4x \,\,\square
\sqrt{-g}\bigl(g^{\mu\nu}\bar\ph_\mu\ph_\nu +
\onetwo\,\bar\ph\ph -
\onetwo\, \bar\ph b\bigr)
\cr}\eqno(2.3)
$$
The limit of $r^{1/2}\ph(r,t,\Omega)$, as $r$ tends to infinity, is the
boundary field $\tilde \ph(t,\Omega)$, while
the others fall off as $r^{-5/2}$ and make no contribution to the boundary
theory. Gauge
transformations yield to the BRST transformation
$$
\delta (\ph,b,c,d) = (c,0,0,b).\eqno(2.4)
$$

It is easy to solve the field equations, and the result is that the
solution space carries the
non-decomposable representation
$$
D[\ph] := D(\fivetwo,0) \to D(\onetwo,0) \to D(\fivetwo,0)\eqno(2.5)
$$
of the anti-De Sitter group. The physical modes, in $D(\onetwo,0)$, are
distinguished from the others
in that they fall off slowly, as $r^{-1/2}$, at spatial infinity. The free
field $b$ is
identified with the invariant subspace associated with the representation
on the right; the free
fields $c$ and $d$ transform the same way and $b,c,d$ all fall off as
$r^{-5/2}$ at infinity.

The canonical vector current is closed, and exact up to the contribution of
the boundary
term, which testifies to the fact that free singletons do not contribute to
the charge
of the bulk. It is conserved, and therefore it cannot be identified with
the vector potential.

Instead, we consider the current
$$
\hat j^\mu[\ph] = \sqrt{-g}g^{\mu\nu}(\bar\varphi \varphi_\nu -
 \bar \varphi_\nu\varphi) := \sqrt{-g}g^{\mu\nu}\hat j_\mu,\eqno(2.6)
$$
and the vector potential
$$
A_\mu = e\, \hat j_\mu[\ph] = e\,(\bar\ph \varphi_\nu -
 \bar\ph_\nu\varphi).\eqno(2.7)
$$

The action of $SO(3,2)$ on the constituent field $\ph$ induces an action on
the composite field $A$.
When the field $\ph$ is free, then so is $A$, in the sense that the induced
representation is now
contained in the direct product
$$
D[\ph] \otimes D[\ph],\eqno(2.8)
$$
that is equivalent to a direct sum of massless representations [3].
Because the fields are multiplied locally, only a small part of the direct
product is carried by
the field, namely the non-decomposable representation
$$
D[A] := D(3,0) \to D(2,1) \to D(3,2).\eqno(2.9)
$$
This is precisely a Gupta-Bleuler triplet of anti-De Sitter electrodynamics
[8].

It has been shown [6] that the BRST transformation of $\ph$ induces the
usual BRST transformation of
$A$. Therefore the Lorentz condition, $b = 0$, of the free singleton field
theory, induces the
usual Lorentz condition on $A$; this tallies perfectly with the formula
$$
\partial_\mu A^\mu = e\,\bigl(\bar\ph b - \bar b\ph\bigr)\eqno(2.10)
$$
that comes from the definitions of $A$ and $\hat j[\ph]$ and the free field
equations.

When   free  singleton modes are inserted for $\ph$ and for $\bar
\ph$ in the expression for $A$, then a physical massless mode (transverse
polarization) is produced.
If one of the two factors in the product is physical and the other is a
gauge mode, then the field
mode
$A$ that results is a gauge mode; in other words, it is a gradient.
Therefore, if the potential is
coupled to a conserved current, then free singleton gauge modes decouple,
so long as both factors in
$A$ are free fields. This gives some encouragement for hoping that, under
the right circumstances,
such a coupling may give a unitary field theory.

Precisely, our proposal is  as follows. Instead of the ordinary
electromagnetic potential, couple
singleton fields to any conserved current $J$ by introducing the interaction
$$
\int d^4x J^\mu (x)A_\mu(x),\eqno(2.11)
$$
with
$$
A_\mu(x) = e\,\hat j_\mu(x).\eqno(2.12)
$$
The physical singleton Fock space includes massless particles with all integer
spins, as 2-singleton composites, but only two-singleton states with the
quantum numbers of photons
couple directly to the vector current $J$ of ordinary particles.
One-particle singleton states are extraordinarily hard to detect,
for  kinematical reasons if not in principle [3].

There may be a singleton contribution to $J$, but it has to be conserved,
so that $J$ may include the canonical current $j[\ph,b]$ (that vanishes on
the physical subspace), but not $\hat j[\ph]$.
  \b
\no{\it Composite gravity.}

To compose gravitons out of singletons is just as natural. However, the
canonical, bulk, energy
momentum tensor cannot be identified with the gravitational potential since
it is conserved.
The complete expression for it is, in the case of a real singleton field,
$$
t_{\mu\nu}[\ph,b] = \onetwo \bigl(\ph_\mu b_\nu + \ph_\nu b_\mu  -
g_{\mu\nu}(g^{\sigma\tau}\ph_\sigma b_\tau - u \,\ph b)\bigr)
+ \delta(\infty)\bigl(\tilde\ph_i\tilde \ph_j -
\onetwo \tilde g_{ij}\tilde g^{kl}\tilde \ph_k\tilde\ph_l),\eqno(2.13)
$$
where a tilde indicates boundary values. Except for the boundary term it is
BRST-exact; the
physical part is thus concentrated on the boundary. The tensor $\hat
t\,[\ph]$ that is needed is
$$
\hat t_{\mu\nu}[\ph] = \ph_\mu \ph_\nu - \onetwo(g^{\sigma\tau}\ph_\sigma
\ph_\tau - u \ph
\ph),\eqno(2.14)
$$
it satisfies, by virtue of the free field equations,
$$
\partial_\mu \hat t^{\mu\nu}[\ph] = \ph_\nu b.\eqno(2.15)
$$
This too vanishes on the physical subspace, and it makes sense to identify
the tensor $\hat t\,[\ph]$ with the gravitational potential,
$$
h_{\mu\nu} = \kappa \,  \hat t_{\mu\nu}[\ph].\eqno(2.16)
$$

When both factors in the field product (2.14) are replaced by free field
modes, then the action
of $SO(3,2)$ on the tensor current, and on the gravitational potential, is
reduced to
$$
D(4,1) \to D(3,2) \to D(4,1),\eqno(2.17)
$$
which is precisely the Gupta-Bleuler triplet associated with free anti-De
Sitter gravitons.
We therefore propose a theory of quantum gravity in which the first order
perturbation of the
background anti-De Sitter metric is given by Eq.s (2.16) and (2.14). This
tensor field can be
coupled to any conserved energy-momentum tensor $T$, by introducing the
interaction
$$
\int d^4x \,T^{\mu\nu}(x)h_{\mu\nu}(x).\eqno(2.18)
$$
Just as was explained in connection with the vector potential, it is not
possible to include
$\hat t[\ph]$ as a contribution to $T$; what has to be included in $T$ is
the conserved energy momentum
tensor
$t[\ph,b]$ given in (2.13). Of course, this interaction must be corrected
by nonlinear terms,
in the usual way. In addition, the metric $g$ in (2.14) and in the volume
element must be
replaced by the perturbed metric. The result is that (2.14) and (2.16) will
give, not a closed
expression for the metric in terms of the singleton field, but a nonlinear
relation between
both that can be solved for the metric as a power series in $\ph$, the
leading terms
being $g_{\mu\nu}  + \kappa \hat t_{\mu\nu}$.

 \bb

\no {\steptwo 3. Spinors.}

The spinor singleton field is also governed by a dipole. The complete
Lagrangian is
$$\eqalign{
\L &= \L_4 + \L_3,\cr
\L_4 &= \int d^4x \sqrt{-g}\bigl(\itwo (\bar\psi \,\slash \hskip-2mmD b -
\overline{\slash
\hskip-2mmD\psi} b) + \itwo (\bar b \, \slash \hskip-2mm D\psi -
\overline{\slash \hskip-2mm D
b}\psi) -\bar b b + v(\bar \psi b + \bar b \psi)\bigr),\cr
\L_3 &= \onetwo \int d^4 x\sqrt{-g}D_\mu(-\bar \psi \gamma^\mu \slash
\hskip-2mm D\psi -
\overline{\slash \hskip-2mm D \psi} \gamma^\mu \psi + i\bar b \gamma^\mu
\psi - i \bar\psi\gamma^\mu
b).
\cr}\eqno(3.1)
$$
The contribution from the Faddeev-Popov ghosts was omitted, see [6].   The
constant $v = \onetwo
\sqrt
\rho$. The Hamiltonian was calculated in [9], and the entire energy
momentum tensor in [10]; it
is BRST exact except for a boundary term. The remarks about the canonical
conserved vector and tensor
currents apply here too. The vector current
$$
\hat j^\mu[\psi] =  \sqrt{-g}\bar \psi \gamma^\mu \psi
$$
satisfies
$$
\partial_\mu \hat j^\mu[\psi] = \bar\psi b - \bar b\psi;
$$
The tensor current
$$
\hat t^{\mu\nu}[\psi] = \itwo\sqrt{-g}g^{\mu\sigma}\bigl( \bar\psi_\sigma
\gamma^\nu\psi - \bar \psi\gamma^\nu\psi_\sigma\bigr)
$$
satisfies
$$
\partial_\mu \hat t^{\mu\nu}[\psi] = i\sqrt{-g}g^{\nu\sigma}\bigl( \bar
\psi_\sigma b - \bar b_\sigma \psi -{\rm conj.}).
$$
The solutions of the field equations carry the triplet
$$
D(2, \onetwo) \to D(1,\onetwo) \to D(2,\onetwo).
$$
When these free modes are inserted into the vector current $\hat j[\psi]$
one finds that $SO(3,2)$
acts by the same representation $D[A]$ as on the vector current $\hat
j[\ph]$. Similarly, the tensor
current transforms by the same representation as $\hat t\,[\ph]$. (A
symmetrized form of $\hat t\,[\psi]$ must be used.)

This makes it possible to identify the vector potential with $\hat j[\ph]$
or with $\hat
j [\psi]$
or with the sum of both; the same remark applies to the tensor current. For
reasons having to do with
the counting of states in flat limit (difficult) it is likely that the sum
of both is the correct
choice.

\bb

\no{\stepone 4. Supersymmetry.}

The scalar and spinor singleton representations can be combined to a
representation of the superalgebra
$OSp(4/1)$. A superfield formulation has been given, including a
constraint-free Lagrangian
formulation [11]. A scalar superfield that contains both fields has the form
$$
\Phi = \ph + \theta \,\psi + \itwo \theta \gamma^\mu\,\theta\, A_\mu + ...~.
$$
The field equations are
$$
(\square_s - 3)\Phi = B,~~ (\square_s - 3) B = 0,
$$
where $\square _s$ is a super d'Alembertian and $B$ is the
Nakanishi-Lautrup superfield. The
Lagrangian has the same structure as in (2.3) and (3.1), $\L = \L_4 + \L_3$
with
$$
\L_4 = \int d^4x d^4\theta \sqrt{-g}\bigl(-\itwo(Q^\alpha\Phi)(Q_\alpha B)
+ 3\Phi B + \onetwo BB
\bigr).
$$
The operators $Q_\alpha$ are the components of a supercovariant derivative.
In addition to composite photons and gravitons, one can now construct
neutrinos and gravitinos.
The vector super current
$$
\hat j_\alpha[\Phi] = \Phi Q_\alpha \Phi
$$
satisfies
$$
Q_\alpha \hat j^\alpha =  \Phi B
$$
and can be identified with the Wess-Zumino spinor super potential, in the
De Sitter formulation.

In addition to the vector and tensor currents one can construct a spinorial
current that may serve
to construct a composite gravitino, but it is much more attractive to build
the entire
supergravity multiplet of fields from a super stress tensor, or rather from
a tensor that is
similar to the super stress tensor in the way that the vector and tensor
currents $\hat j$ and $\hat t$ are
patterned on the canonical current and stress energy tensors.
We propose to use the tensor
$$
\hat t_{\alpha\beta}[\Phi] = (Q_\alpha\Phi)(Q_\beta \Phi) + \itwo
(Q^\alpha\Phi)(Q_\alpha
\Phi) - 3\Phi\Phi;
$$
it satisfies
$$
Q^\alpha \hat t_{\alpha\beta}[\Phi] = B Q_\beta\Phi,
$$
and it is thus conserved on the physical subspace. The details of a
formulation of linear
anti-De Sitter supergravity, in terms of this type of superfield, remains
to be worked out.
Nevertheless, it is clear that this super tensor carries the right degrees
of freedom to be
identified as a super gravity potential.

\bb

\no{\stepone 5. The problem of unitarity.}

Here we present some additional considerations, going beyond the arguments
given in the introduction,
concerning the question of whether or not the construction proposed in this
paper may, somewhat
miraculously, lead to a unitary theory of composite massless particles and
fields.

We consider some of the simplest Feynman diagrams, involving a  vector
current $J$ made up
of ordinary fields, and a number of lines that represent singletons. If all
of the latter are
external, then we know that there is no problem, for when a pair of free
singletons couple to a
conserved current then only the  physical 2-singleton modes are effective.
Now let us look at a
diagram that has at least one internal singleton line, and two external
singleton lines extending
from vertices located at $x$ and at $x'$. So we are dealing with the
following object,
$$
\ph(x)\ph(x')K(x,x')
$$
and similar quantities containing a derivative of one or the other (or
both) of the $\ph$'s.
The fields $\ph(x)$ and $\ph(x')$ are thus free. Consider an operator
product expansion,
$$
\ph(x)\ph(x') = \sum f_n(x-x')^{(n)}\ph(x)\partial_{(n)}\ph(x),
$$
where $\partial_{(n)}$ is an $n$'th order differential operator.
The field $\ph(x)\partial_{(n)}\ph(x)$ is a massless field with spin $n$.
There is nothing
unphysical about massless composite states with arbitrary spin, provided
that they couple in such
a way that only physical states interact.  But this poses very strong
conditions on the factor
$K(x,x')$.  Only string theory can boast of miracles of this type.

Nevertheless, one solution to this problem is already known. It was shown,
in fact, that by adopting
unconventional field quantization for the singleton fields, it can be
arranged that the composite
operator $\hat j[\ph](x)$ satisfy exactly the canonical commutation
relations of a standard, massless
vector potential. This merely demonstrates the existence of a solution, we
do not wish to claim
that it is the only solution, and in fact, we suggest that it may not be
the best one. To advance
the discussion, we shall explain why we think that the old solution may not
be perfect.

First, there is some sense of diappointment in discovering that this old
``composite electrodynamics"
is precisely equivalent to ordinary electrodynamics. True, the latter does
not need improvement,
but the same is not true of quantum gravity. The type of softening of
interactions at small distances that
is expected to be the most important result of replacing elementary fields,
in this case the gravitational potential,
with composites, is highly desirable in a theory that is non-renormalizable
and hence internally inconsistent. Concerning
quantum gravity there were two problems. First, there is little point in
setting up an elaborate structure of composite gravitons
if there is no gain relatively to the problematic naive version. The second
problem was more technical. If we
apply our old alternate quantization paradigm to a theory with just one
real, scalar singleton
field, then we do not get any interaction at all, simply because the tensor
$\hat t[\ph]$ is symmetric
in the two factors. We conclude, from the discussion of Feynman diagrams,
that conventional
quantization is unlikely to work out, and from this paragraph that our old
solution to the problem, though successful when applied
to QED, is also somewhat unpromising, at least for quantum gravity.

Recall that Wigner [12] questioned conventional quantization of elementary
variables on the grounds
that the most important observables are second order polynomials. He
therefore proposed to examine
commutation relations (among the basic observables, the singletons in our
case) that give
reasonable commutation relations for these observables (the massless
fields). This is very
appropriate in our context, since isolated singletons are essentially
unobservable, if not
in principle, then for kinematical reasons [3]. Wigner's suggestion led to
the discovery of
parastatistics. (Later, it was proposed   to apply parastatistics to quarks
[13], but this
idea was subsequently transformed into the popular concept of color [14].)
And this is the only
alternative to strict canonical quantization that has been attempted (and
not very dilligently) in
quantum  field theory.

Clearly, the whole difficulty with unitarity can be ``solved'' by
postulating free field
commutation
relations for the composite operators. But this would be to beg the
question. What is needed is to
take a  good look at what are the possibilities available for the
quantization of singleton fields, since, like quarks, they do not
need to be represented by local field operators. It is not
unlikely that   string theory may provide further inspiration for attacking
this problem.

In this paper we have shown that massless fields may be regarded as
composites, and that the constituents have to be gauge fields,
hence singletons. On a certain level this works in an easy and natural manner.
The problem of unitary is a serious one, and well worth investigation.

\ve
\no{\stepone Acknowledgements.}

This paper was inspired by a stimulating collaboration with Sergio Ferrara.
\bb

\no{\stepone References.}

\item {[1]} J. Maldacena, hep-th/9711200. K.Stefsos and K. Skenderis,
hep-th/9711138. P. Claus, R.
Kallosh and A. van Proeyen, hep-th/9711161. P. Claus, R. Kallosh, J. Kumar,
P. Townsend and A. van
Proeyen, hep-th/9801206. S. Ferrara and C. Fr\o nsdal, hep-th/9712239;
hep-th/9802126.
E. Witten, hep-th/9802150.

\item{[2]}S. Ferrara, C. Fr\o nsdal  and A. Zaffaroni, hep-th/9802203

\item {[3]} M. Flato and C. Fronsdal, Lett. Math. Phys. {\bf 2} (1978) 421;
Phys.Lett. {\bf 97B},
(1980) 236; J. Geom. Phys. {\bf 6} (1988) 294.

\item {[4]} -------------, Phys.Lett. {\bf 172} (1986) 412.

\item {[5]} -------------, M. Flato and C. Fronsdal, J. Math. Phys. {\bf
22} (1981) 1100.

\item {[6]} -------------, Phys. Lett. {\bf 189} (1987) 145.

\item {[7]} -------------,  Comm. Math. Phys. {\bf 108} (1987) 469.

\item {[8]} C. Fronsdal Phys. Rev. D {\bf 12} (1975) 3810; B. Binegar, C.
Fronsdal and W. Heidenreich, Ann. Phys. {\bf 149} (1983)
254.

\item {[9]} U. Percoco, Lett. Math. Phys. {\bf 12} (1986) 469.

\item {[10]}  F. Yat-lo Lai, PhD Thesis, UCLA 1997.

\item {[11]} C. Fronsdal, Lett. Math. Phys. {\bf 16} (1988) 163; {\it idem}
page 173.

\item{[12]} E.P. Wigner, Phys. Rev. {\bf 77} (1950) 711.

\item{[13]} O. W. Greenberg and A.M.L. Messiah, Phys. Rev. {\bf 138B}
(1965) 1155.

\item{[14]} M. Gell-Mann, Proc. 13'th Int. Conf. High-Energy Physics,
Berkeley 1966.

\end